\documentclass[12pt]{iopart}
\expandafter\let\csname equation*\endcsname\relax
\expandafter\let\csname endequation*\endcsname\relax
\usepackage{amstext}
\usepackage{amssymb}
\usepackage{graphicx}
\usepackage{dcolumn}
\usepackage{bm}
\usepackage{subfigure}
\usepackage{xcolor}
\usepackage{cite}
\usepackage{amsfonts}
\usepackage{amssymb}
\usepackage{amsmath}
\usepackage{txfonts}
\usepackage{bm}
\usepackage{dsfont}
\usepackage{tikz}
\usepackage{graphicx}

\begin{document}

\title{Quantum Dynamical Resource Theory  under  Resource Non-increasing  Framework}
\author{Si-ren Yang$^1$, Chang-shui Yu*$^{1,2}$}
\begin{abstract}
We define the resource non-increasing (RNI) framework to study the dynamical resource theory.  With such a definition, we propose several potential quantification candidates under various free operation sets. For explicit demonstrations, we quantify the quantum dynamical coherence in the scenarios with and without post-selective measurements. Correspondingly, we show that maximally incoherent operations (MIO) and incoherent operations (IO) in the static coherence resource theory are free in the sense of dynamical coherence.  We also provide operational meanings for the measures by the quantum discrimination tasks. Moreover, for the dynamical total coherence, we also present convenient measures and give the analytic calculation for the amplitude damping channel. \end{abstract}
\address{$^1$School of Physics, Dalian University of Technology, Dalian 116024,
China}
\address{$^2$DUT-BSU joint institute, Dalian University of Technology, Dalian 116024, China }
\ead{ycs@dlut.edu.cn} \vspace{10pt}
\begin{indented}
\item[\today]
\end{indented}

\maketitle
\section{Introduction}

In recent years, quantum features, such as quantum entanglement  \cite{entanglement, entangle,e5} and quantum discord \cite{discord,qd1,qd2,qd3,qd4} are taken regarded as quantum resources and quantitatively investigated under rigorous mathematical methods, i.e. quantum resource theories (QRTs), which has been systematically developed since quantum coherence was formally and quantitatively studied in Ref. \cite{Quantifying}. QRTs are powerful approaches to investigate quantum characteristics and have deep effects on our recognization of quantum science. 
 Up to now, QRTs have been widely applied to other quantum features:  nonlocality \cite{nonlocality}, contextuality \cite{contextuality}, non-Gaussianity \cite{Strobel424} and asymmetry \cite{as} and so on. 
QRTs are usually defined by two fundamental ingredients: free states and free quantum operations.  Free states are those states without any resource and free quantum operations are referred to as those that could not generate quantum resources if operated on free states. 
In such a framework, QRTs bring new insight in studying quantum features on the static level: measures for quantum features not only evaluate the weight of a system but also have operational meanings corresponding to specific quantum processes. Such a framework for static resources is quite rigorous, and could even develop a connection to different areas. For example, in the QRTs of quantum coherence, researchers proposed multiple measures \cite{alter,trace,multi,mea1,mea2,mre,mea3} and  operational interpretations \cite{op1,op2,op3,op4,op6}. It has been shown that quantum coherence  closely correlated with other quantum features like quantum entanglement \cite{ entangle,e1,e5}, quantum discord \cite{discord,qd1,qd2,qd3,qd4}, quantum asymmetric \cite{ass1,piani,as3} et al.

Quantum states might get evolved under dynamic processes or quantum operations.  Typically, quantum states can be viewed as some special quantum channels. Most quantum processes can be characterized by quantum channels which are mathematically the set of completely positive trace preserving (CPTP) maps. Corresponding to static states, quantum channels are dynamical quantum resources in quantum science and carry much more information than static systems. Therefore,  one natural idea would be whether to investigate dynamical channels by QRTs. In this sense, one possible way is to follow and upgrade the two elements in static QRTs to dynamical levels, and find out what could be gotten in dynamical resources through the QRTs.  With such motivations, researchers are working on this area and have made progresses \cite{ch3,ch5,POVM1,qrt5,ch9,ch8,ch1,qrt1,nch1,nch2,nch3,nch4,qd,nch5,nch6}. Some studied the dynamical coherence via different free channels, for example, the  Choi isomorphism of classical channels is considered \cite{ch3}. In Ref. \cite{POVM1}, the resource theory proposed free Positive-Operator-Valued Measures (POVMs) and free detection/creation incoherence in the sense of quantum computational setting \cite{qrt5}.  In addition, the dynamical entanglement has been considered in Ref. \cite{ch8}, and a quantitative relationship between the dynamical coherence and the dynamical entanglement are introduced in Ref. \cite{ch1}.

In dynamical QRTs, the two ingredients are free operations and free superoperations (completely CP and TP preserving maps). The free operations are those without expected quantum features, and free superoperations are defined as those that could not map a free operation to a resourceful operation. However, one would find that the definitions of free superoperation sets depend on the physical considerations. Even for the same quantum feature, the free sets are not unique. 
Another tough problem is how to measure dynamical quantum features. Due to the diversity of channel representations,  the analytical solutions rarely exist,  thus measures are usually given in numerical results. Moreover, realizations of superoperations are not unique either. To sum up,  dynamical QRTs are developing in different directions, which leads to various QRTs. 

In this paper, we direct a new path to define the free operation sets for quantum dynamical resources. We call it resource non-increasing (RNI) frameworks, in which the free channel won't increase the resourcefulness of the input states.    It will be shown that our RNI mindset provides a straight comprehension for dynamical free and meanwhile guarantee RNI framework has no conflict with the well-defined RNG framework. We also refer to static resource theories and design appropriate superoperations for dynamical QRTs. The free dynamical sets in the RNI frameworks are fairly pellucid.  We present several potential quantifications of dynamical resources under different free operation sets. To investigate the dynamical coherence, we demonstrate that maximally incoherent operations (MIO) and incoherent operations (IO) in the static coherence resource theory are free in the sense of dynamical coherence. In this sense, we give the corresponding measure of the quantum dynamical coherence in the case with and without post-selective measurements. Semidefinite programming (SDP) is also applied to quantify dynamical coherence without post selective measurements. In addition, we also study the quantification of the dynamical total coherence, for which an analytic calculation is given for the amplitude dumpling channel.   We organize the remaining parts of this paper as follows. In Sec. II, we review the dynamical QRTs, propose our RNI framework and present several alternative quantifications. In Sec. III, we establish dynamical QRTs for quantum coherence in different scenarios. We illustrate the operational meanings of measures in quantum discrimination tasks.  In Sec. IV,  we investigate the dynamical total coherence and give an analytic calculation as an example. We summarize the paper in Sec. V.

\section{Resource theory of quantum channels}
RNI framework has a direct definition of the free dynamical set, i.e. the free operations could not increase any quantum static resource for arbitrary static input.  Therefore, to demonstrate the RNI frameworks, we need the unambiguous definitions of static resource measures first. As mentioned previously, for static resource theory, the free states are those without any resource, and the set of free states are denoted by $\mathbb{F}$. The free operations with its Kraus operators $\{K_i\}$ are defined by $K_i\delta K_i^\dagger\in \mathbb{F}$ for any free state $\delta$. Thus a valid static resource measure can be given as follows. 

\textit{Proposition 1.-} A static resource measure $\mathcal{M}$ for some certain quantum resource $\mathcal{R}$  should fulfill (i) Faithfulness:  $\mathcal{M}\geq 0$ and vanishes for free states; (iia) Strong monotonicity:  the average resource under selective free operations can not be increased; (iib) Monotonicity:  the resource of the state after free operations can not be increased; (iii) Convexity: mixing states would not increase their resourcefulness. These constraints are widely applied in measuring entanglement, coherence et al \cite{Quantifying}.
 
 It is known to all that the strong monotonicity combined with the convexity would lead to the general monotonicity. The measures that fulfill the strong monotonicity imply that such quantum channels are post-selective measurements allowed. These selective measurements exist in a scenario that the results are accessible with post-selective operations. With the static resource measure, we can present our free operations in the RNI framework (RNI-free operations).  
 
 \textit{Theorem. 2-} RNI-free operations in the dynamical resource theory are consistent with the free operations in the static resource theory.
 
 \textit{Proof.-} Based on the idea of the RNI framework, the RNI-free operations $\mathcal{E}(\cdot)=\sum_{i}K_{i}\cdot K^{\dagger}_{i}$ with respect to a valid static resource measure $\mathcal{M}$ is defined as $\sum_n \mathrm{Tr}((K_{i}\rho K^{\dagger}_{i}))\mathcal{M}(K_{i}\rho K^{\dagger}_{i}/\mathrm{Tr}((K_{i}\rho K^{\dagger}_{i}))\leq \mathcal{M}(\rho)$ for any density matrix $\rho$. If $\rho\in \mathbb{F}$, due to the faithfulness of static measures, $\mathcal{M}(\rho)=0$, so $\mathcal{M}(K_{i}\rho K^{\dagger}_{i}/\mathrm{Tr}((K_{i}\rho K^{\dagger}_{i}))=0$, which is consistent with the definition of free operations in the static resource theory. On the contrary, if $\mathcal{E}(\cdot)=\sum_{i}K_{i}\cdot K^{\dagger}_{i}$ is the static free operations, based on the strong monotonicity, the mentioned definition for the RNI-free operations is also satisfied. \hfill{}$\Box$
 
Note that if a quantum channel is considered in a black-box scenario, which means the measurement is non-selective, one can only require monotonicity in a non-increasing framework, i.e. $\mathcal{M}(\mathcal{E}(\rho))\leq \mathcal{M}(\rho)$. This corresponds to the RNI framework in the sense of monotonicity. Similarly, the free operations can also be naturally defined subject to monotonicity, which will be directly used later and won't be explicitly elucidated here.
 
 Another ingredient in dynamical QRTs is the free superoperations. Similar to the static resource theory, the free superoperation has a primal constraint that maps a free operation to free operation. Besides, we stress that the construction of superoperations should not contain certain resourceful ingredients. Now we propose our free superoperations with the following structure.
 
\textit{Definition. 3-} A superoperation  represented by Kraus operators $\{ \mathfrak{F}_n\}$ is free if and only if $\mathfrak{F}_n[\mathcal{N}]$ can be written into the sequence as
 \begin{equation} 
 \mathfrak{F}_n[\mathcal{N}]=\mathcal{E}_{i_n,\mathrm{\Phi}^{j_n}_n},\cdots,\mathcal{E}_{i_2,\Phi^{j_2}_2}\mathcal{E} _{i_1,\mathrm{\Phi}^{j_1}_1} \label{superchannel},
 \end{equation} 
 where $\mathcal{E}_{i_n,\mathrm{\Phi}^{j_n}_n}$ denotes the $j_n$th Kraus element of the superoperation $\{\mathcal{E}_{i_n,\mathrm{\Phi}^{j_n}_n}\}$ with the corresponding free operation $\{\mathrm{\Phi}^{j_n}_n\}$ and 
 \begin{align}
 & \mathcal{E}_{i_n=0,\mathrm{\Phi}^{j_n}_n}[\mathcal{N}]=\mathrm{Tr}[\mathcal{N}],\notag\\
 &\mathcal{E}_{i_n=1,\mathrm{\Phi}^{j_n}_n}[\mathcal{N}]=\mathrm{\Phi}^{j_n}_n\circ \mathcal{N},\mathcal{E}_{i_n=2,\mathrm{\Phi}^{j_n}_n}[\mathcal{N}]=\mathrm{\Phi}^{j_n}_n\otimes \mathcal{N}, \notag\\
 &\mathcal{E}_{i_n=3,\mathrm{\Phi}^{j_n}_n}[\mathcal{N}]=\mathcal{N}\circ\mathrm{\Phi}^{j_n}_n,\mathcal{E}_{i_n=4,\mathrm{\Phi}^{j_n}_n}[\mathcal{N}]=\mathcal{N}\otimes\mathrm{\Phi}^{j_n}_n.\label{superele}
 \end{align}
 
The above definition implies the tensor product structure is automatically satisfied if one replaces $\mathrm{\Phi}^{j_n}_n$ by identity operator $\mathds{1}$. Furthermore,  it is not difficult to find that our superoperations would be free for every single Kraus operator. Combing the two characteristics one can conclude that our superoperations are separately (every Kraus operator free) and completely (tensor product structure) free.  Hence, our free superoperations meet the requirement of our motivation.  With the two ingredients defined, we can define the measure of RNI dynamical resource as follows. 
 
 \textit{Definition 4.-} A  measure $T(\cdot)$ quantifying dynamical resourcefulness of arbitrary quantum channel  $\mathcal{N}$  is a qualified measure if the following are satisfied.
 \begin{align}\nonumber
(1)&\text{ Faithfulness}: T(\mathcal{N}) \geq 0, \text{where equality holds iff}\  \mathcal{N} \text{ is free};\\ \nonumber
(2)&\text{ Monotonicity or strong monotonicity}: T(\mathcal{N}) \geq T(\mathfrak{F}[\mathcal{N}]) \\ \nonumber 
&\text{or} \ T(\mathcal{N})\geq\sum_{n}p_{n}T(\mathcal{N}_{n}) \text{ for free superoperations} \\ \nonumber & \mathfrak{F}=\sum_{n}p_{n}\mathfrak{F}_{n} \text{ with  } 
 \mathcal{N}_{n}=\mathfrak{F}_{n}[\mathcal{N}]\: \text{and}  \sum_n p_n=1,\\\nonumber
(3)&\text{ Convexity}:  T(\mathcal{N}) \text{ is convex}.\nonumber
 \end{align}

The strong monotonicity is an operational constraint where the measures obey the monotonicity under selective measurements. Our free superoperations allow us to read out information of the post states.
However, if one could not have enough information about the resourceful superoperation (such as a black box in some practical scenarios mentioned previously),  it is enough to consider the monotonicity, even though in a QRT strong monotonicity is required. Our free superoperations can also be explicitly given in the sense of quantum computational settings, which is similar to Ref.  \cite{qo}.   

Up to now, we have well established the fundamental requirements for a dynamical resource measure in the RNI framework. Thus in the following, we present two natural quantification approaches, one is based on the distance from the free operation set, the other is directly based on the violation of the definition of free operations.

 \textit{Definition. 5-} Dynamical resource of  the channel $\mathcal{N}(\cdot)=\sum_{i}K_{i}\cdot K^{\dagger}_{i}$  can be measured by  the minimal distance from the free set $\mathcal{S}$ with appropriate distance functions  $\vert\vert \cdot\vert\vert_{\star}$ as
 \begin{align}
T(\mathcal{N})&=\displaystyle \min_{\mathcal{F}\in  \mathcal{S}} \vert\vert \mathcal{N} -\mathcal{F}\vert\vert_{\star}, \label{normmeasure}
\end{align}
or by the magnitude of the violation  of the free operation as
\begin{align}
\tilde{T}(\mathcal{N})&=\displaystyle \max \{\mathrm{\Delta} \mathcal{M}_{\star}(\mathcal{N}), 0\},\label{minus}
\end{align}
with $\mathrm{\Delta} \mathcal{M}_{\star}(\mathcal{N})=\max_\rho \sum_i\mathrm{Tr}[K_i\rho K_i^\dagger]\mathcal{M}\left(K_i\rho K_i^\dagger/\mathrm{Tr}[K_i\rho K_i^\dagger\right)$ where $\star$ denotes the proper distance functions in Eq. (\ref{normmeasure})  or static resource measure in Eq. (\ref{minus}) such that both $T(\mathcal{N})$ or $\tilde{T}(\mathcal{N})$ satisfies definition 4.

 Finally, we'd like to emphasize that the dynamical resource in the RNI framework makes sense for the quantum channel with given Kraus operators since our definitions are based on the strong monotonicity. Of course, one can consider the channel without post-selection in the sense of monotonicity. 

 \section{Dynamical  quantum coherence in resource non-increasing framework}

Now we will consider the dynamical resource theory of coherence in the RNI framework. Since we have shown that the RNI-free operations are the same as free operations in the static resource theory, the RNI-incoherent operations in the dynamical resource theory naturally correspond to the incoherent operations in the static coherence quantification. As mentioned in the previous section, the RNI dynamical resource theory can be considered in the sense of both strong monotonicity and monotonicity. One can find that the free operations subject to the dynamical coherence are separately the incoherent operations (IO) and the maximally incoherent operations (MIO). Thus, we will directly employ the proposed approaches to quantify the dynamical coherence based on definition 5.

 \textit{Theorem 6.-} Let the free operation set denote by IO and MIO corresponding to the RNI coherence with and without post-selection scenarios,  the dynamical coherence for a quantum channel $\mathcal{N}$ can be measured by
\begin{align}
T_{1/\diamond}(\mathcal{N})=\displaystyle \min_{\mathcal{F}\in\text{IO/MIO}} \vert\vert\mathcal{N}-\mathcal{F}\vert\vert_{1/\diamond},
\end{align}
where the induced trace norm and the diamond norm are defined as
\begin{align}
\vert\vert\mathrm{\Phi}\vert\vert_{1}=\text{max}\{{\vert\vert\mathrm{\Phi(X)}\vert\vert_{1}:\;X\in\mathcal{L}(X),\: \vert\vert X\vert\vert_{1}\leq 1}\},
\end{align} 
\begin{align}
\vert\vert\mathrm{\Phi}^{\mathrm{A}\rightarrow\mathrm{B}}\vert\vert_{\diamond}=\vert\vert\mathrm{\Phi}^{\mathrm{A}\rightarrow\mathrm{B}}\otimes\mathds{1}^{\mathrm{C}}\vert\vert_{1}.
\end{align}

\textit{Proof.-} Firstly,  $T_{1/\diamond}(\mathcal{N})$ vanishes for free operations since it is defined by the minimal distance. For the convexity, let's consider a channel $\mathcal{N}$ mixed by a set $\mathcal{N}_m$ with probabilities $q_m$ and  denote the corresponding optimal free operation in
 set IO/MIO by $\mathcal{F}_m$, the  average coherence is
\begin{align}
&\sum_{m} q_{m} T_{1/\diamond}(\mathcal{N}_m)=\sum_m q_m\vert\vert\mathcal{N}_m-\mathcal{F}_m\vert\vert_{1/\diamond}\nonumber\\
&\geq \vert\vert\sum_m q_m\mathcal{N}_m-\sum_m q_m \mathcal{F}_m\vert\vert_{1/\diamond}
\geq \vert\vert\mathcal{N-F}\vert\vert_{1/\diamond}\nonumber\\
&=T_{1/\diamond}(\mathcal{N}).
\end{align}
For the strong monotonicity, let's consider the free superchannel $\mathfrak{F}=\sum_m p_m \mathfrak{F_m}$ given in Eq.(\ref{superchannel}). One can denote $T_{1/\diamond}(\mathfrak{F}_{m}[\mathcal{N}])=T_{1/\diamond}({\mathcal{N}_m})$. Since the induced trace norm is sub-multiplicative and sub-multiplicative with respect to tensor prouduct, which indicates that  $T_{1/\diamond}({\mathcal{N}_m})\leq T_{1/\diamond}(\mathcal{N})$ for every single Kraus operatore in  free superchannels. Thus, the strong monotinicity holds by the following inequality
\begin{align}
\sum_m p_m T_{1/\diamond}({\mathcal{N}_m}) \leq \sum_m p_m T_{1/\diamond}(\mathcal{N})=T_{1/\diamond}(\mathcal{N}).
\end{align}
Finally one can directly obtain the monotonicity based on the strong monotonicity and convexity.

Let $T_{1/\diamond,non}$  denotes $T_{1/\diamond}$ without post-selective measurements. It is shown that $T_{1/\diamond,non}$  have a direct operational meaning in the quantum channels discrimination task and can be calculated by semidefinite programming (SDP).  We will illustrate the details in the appendix. 

Besides the distance measures, one can also employ the maximal violation in definition 5 to define the dynamical coherence as follows. 

  \textit{Theorem 7.-}  Given a quantum channel $\mathcal{N}(\cdot)=\sum_{n}K_{n}\cdot K^{\dagger}_{n}$, the dynamical coherence can be well quantified by
 \begin{align}
\tilde{T}(\mathcal{N})&=\displaystyle \max \{\mathrm{\Delta} \mathcal{M}_{np/p}(\mathcal{N}), 0\},\label{minus}
\end{align}
where  $\mathcal{D}$ denotes the set of all density matrices in the space, and 
\begin{equation} 
\mathrm{\Delta}\mathcal{M}_{np}(\mathcal{N})=\displaystyle \max_{\rho\in\mathcal{D}}\mathcal{C}(\mathcal{N}(\rho))-\mathcal{C}(\rho)\label{spower}
\end{equation}
without post-selective measurements, 
\begin{align}
\mathrm{\Delta} \mathcal{M}_{p}(\mathcal{N})= \max_{\rho\in\mathcal{D}}\sum_n p_n \mathcal{C}(\rho_n)-\mathcal{C}(\rho).
\end{align}
with post-selective measurements. Here $p_n=\mathrm{Tr}((K_{n}\rho K^{\dagger}_{n}))$ and $\rho_n=K_{n}\rho K^{\dagger}_{n}/p_n$.

\textit{Proof.-}
At first, the definition of free operations in either scenario directly implies  $\tilde{T}(\mathcal{N})\geq 0$ which is saturated if and only if  $\mathcal{N}\subset\text{IO/MIO}$. 

For the strong monotonicity, one will have to consider definition 3 in detail, which shows free superoperations can be written as $\mathfrak{F}(\mathcal{N})=\sum_m q_m \mathfrak{F}_m(\mathcal{N})=\sum_m q_m\mathcal{N}_m$. From the following, one can find that every $\mathfrak{F}_m$ implies $\tilde{T}(\mathcal{N}_m)\leq \tilde{T}(\mathcal{N}$), which immmediately leads to $  \sum_m p_m \tilde{T}({\mathcal{N}_m}) \leq \sum_m p_m\tilde{ T}(\mathcal{N})=\tilde{T}(\mathcal{N})$.  (i) Discarding the system with $\mathcal{E}_{i_n=0}$ makes a free state; (ii)  Attatching ancilla by $\mathcal{E}_{i_n=2, 4}$ means $\mathcal{N}(\rho)=\mathrm{Tr_{A}}[(\mathrm{\Theta}_A\otimes\mathcal{N})(\sigma_A \otimes \rho)]=\mathrm{Tr_{A}}[\mathrm{\Theta}_A(\sigma_A)\otimes \mathcal{N}(\rho)]=\mathrm{1}\cdot\mathcal{N}(\rho)$ ; (iii) Linking a free operation by $\mathcal{E}_{i_n=1}$ corresponds to  $\mathcal{C}(\mathrm{\Theta}\circ\mathcal{N}(\rho))\leq\mathcal{C}(\mathcal{N}(\rho))$ for any state $\rho\in\mathcal{D}$ and any free channel $\mathrm{\Theta}$ due to the monotonicity of static measures $\mathcal{C}$; (iv) Linking a free operation for $\mathcal{E}_{i_n=3}$ can be verified as
\begin{align}
\tilde{T}(\mathcal{N}\circ\mathrm{\Theta})&= \max_{\rho}\sum_n p_n \mathcal{C}(\frac{K_n\rho_{\mathrm{\Theta}} K^{\dagger}_n}{p_n})-\mathcal{C}(\rho)\nonumber\\
&\leq  \max_{\rho}\sum_n p_n \mathcal{C}(\frac{K_n\rho_{\mathrm{\Theta}} K^{\dagger}_n}{p_n})-\mathcal{C}(\rho_{\mathrm{\Theta}})\nonumber\\
&\leq  \max_{\rho}\sum_n p_n \mathcal{C}(\frac{K_n \rho K^{\dagger}_n}{p_n})-\mathcal{C}(\rho)\nonumber\\
&=\tilde{T}(\mathcal{N}),\label{13}
\end{align} 
where $\mathrm{\Theta}$ represents any free channel and $\rho_{\mathrm{\Theta}}=\mathrm{\Theta}(\rho)$ is the post state operated by the free channel. The first inequality comes from  the monotonicity of static measures, i.e., $\mathcal{C}(\rho_{\mathrm{\Theta}})\leq\mathcal{C}(\rho)$ and the second holds since the maximum overall the density matrix spaces is definitely not less than the maximum overall the subspace.  Eq. (\ref{13}) mainly focuses on $\mathrm{\Delta}\mathcal{M}_{p}(\mathcal{N})$. A similar proof can be easily obtained for $\mathrm{\Delta}\mathcal{M}_{np}(\mathcal{N})$ (not given here). So (i)$\sim$ (iv) prove the strong monotonicity.

For the convexity, let's first prove  the scenario with post-seletive measurements. Consider the dynamical coherence of mixing a set of  channels $\{\mathcal{N}_i\}$ with probabilities $\{q_i\}$.   Let a state $\rho$ undergo channel $\mathcal{N}_i$ with its Kraus operator $K_{in}$ , the post-measurement state is denoted by $\sigma_{i,n}=K_{in}\rho K_{in}^\dagger$, then we have 
 \begin{align}
 \tilde{T}(\sum_i q_i\mathcal{N}_i)
 &=\max_{\rho \in \mathcal{D}}\sum_{i,n} q_i \mathrm{Tr}(\sigma_{i,n})\mathcal{C}(\frac{\sigma_{i,n}}{\mathrm{Tr}(\sigma_{i,n})}) -\mathcal{C}(\rho)\label{op1g}\\
& =\sum_{i,n} q_i \mathrm{Tr}(\sigma^{0}_{i,n})\mathcal{C}(\frac{\sigma^0_{i,n}}{\mathrm{Tr}(\sigma^0_{i,n})}) -\mathcal{C}(\rho^0)\\
&\leq \sum_{i,n} q_i \mathrm{Tr}(\sigma^{i}_{i,n})\mathcal{C}(\frac{\sigma^i_{i,n}}{\mathrm{Tr}(\sigma^i_{i,n})}) -\mathcal{C}(\rho^i)\\
& = \sum_{i,n} q_i  \max_{\rho^{i} \in \mathcal{D}} p^i_n \mathcal{C}(\rho^i_n) -\mathcal{C}(\rho^i)=\sum_i q_i \tilde{T}(\mathcal{N}_i),
 \end{align}
 where  the superscript on $\sigma^{0}_{i,n}$ denotes the optimal state of the maximum in the sense of Eq. (\ref{op1g}) and $\sigma^{i}_{i,n}$ is the optimal one for $\mathcal{N_i}$. 
 
In the case without post-selection, for the mixed channel $\sum_i q_i\mathcal{N}_i$, we have 
 \begin{align}
  \tilde{T}(\sum_i q_i\mathcal{N}_i)&=\displaystyle \max_{\rho \in \mathcal{D}}\mathcal{C}(\sum_i q_i \mathcal{N}_i(\rho))-\mathcal{C}(\rho)\\
  &=\mathcal{C}(\sum_i q_i \mathcal{N}_i(\sigma))-\mathcal{C}(\sigma)\label{staconvex}\\
  &\leq\sum_i q_i \mathcal{C}(\mathcal{N}_i(\sigma))-\mathcal{C}(\sigma)\\
 &\leq \sum_i q_i \displaystyle  \max_{\rho_i \in \mathcal{D}}\mathcal{C}(\mathcal{N}_i(\rho_i))-\mathcal{C}(\rho_i)\label{maxoverstate}
 &=\sum_i q_i \tilde{T}(\mathcal{N}_i),
 \end{align}
 where $\sigma$ denotes the optimal state for the mixed channel $\sum_i q_i\mathcal{N}_i$,  and the inequality Eq. (\ref{staconvex}) comes from the convexity for static measures.  
 
 Up to now, we have proved the convexity, which will directly lead to the monotonicity associated with the strong monotonicity.  $\Box$

The dynamical coherence in Eq. (\ref{spower}) has a similar form with cohering power in Ref. \cite{PhysRevA.92.032331}, but our maximum is taken over all the density matrices rather than incoherent states. The MIO set was proposed as the maximal set of free operations in the static QRT,  and here we show that the free set of RNI with the non-selective measurements is exactly the MIO.   The RNI free set subject to  MIO can provide a new operational interpretation to the RNG framework.  In this sense, one can find an alternative dynamical coherence measure assisted  by the dephasing channel $\mathrm{\Delta}(\cdot)=\sum_i \langle i\vert \cdot\vert\ i \rangle\vert i\rangle\langle i \vert$ as follows.

\textit{Theorem 8.-}  Given a quantum channel $\mathcal{N}(\cdot)=\sum_{n}K_{n}\cdot K^{\dagger}_{n}$, the dynamical coherence without post measurements can be quantified by
\begin{align}
T_{a,non}(\mathcal{N})&=\displaystyle \min_{\mathcal{F}\in\text{MIO},\delta \in\mathcal{I}} \vert\vert(\mathcal{N}-\mathcal{F})\delta\vert\vert_{1}\\
&=\displaystyle \min_{\mathcal{F}\in\text{MIO}} \vert\vert(\mathcal{N}-\mathcal{F})\mathrm{\Delta}\vert\vert_{1}.
\end{align}

\textit{Proof.-} (1) The distance functions guarantee $T_{a,non}$ can faithfully detect dynamical coherence. (2) The convexity holds because of  the absolute homogeneity and the triangle inequality. Considering two quantum channels $\mathcal{N}$ and  $\mathcal{M}$ such that $T_{a,non}(\mathcal{N})=\vert\vert \mathcal{N}-\mathcal{F}_1\vert\vert_1$ and  $T_{a,non}(\mathcal{M})=\vert\vert \mathcal{M}-\mathcal{F}_2\vert\vert_1$,  for any $0\leq t \leq 1$, one can find
\begin{align}
T_{a}(t\mathcal{N}+(1-t)\mathcal{M})&=\displaystyle \min_{\mathcal{F}\in \text{MIO}}\vert\vert(t\mathcal{N}+(1-t)\mathcal{M}-\mathcal{F})\mathrm{\Delta}\vert\vert_1\\\nonumber
&\leq\vert\vert(t\mathcal{N}+(1-t)\mathcal{M})\mathrm{\Delta}-(t\mathcal{F}_1+(1-t)\mathcal{F}_2)\mathrm{\Delta}\vert\vert_1\\\nonumber
&=\vert\vert t(\mathcal{N}\mathrm{\Delta}-\mathcal{F}_1\mathrm{\Delta})+(1-t)(\mathcal{M}\mathrm{\Delta}-\mathcal{F}_2\mathrm{\Delta})\vert\vert_1\\\nonumber
&\leq t\vert\vert(\mathcal{N}-\mathcal{F}_1)\mathrm{\Delta}\vert\vert_1+(1-t)\vert\vert(\mathcal{M}-\mathcal{F}_2)\mathrm{\Delta}\vert\vert_1\\\nonumber
&=tT_{a}(\mathcal{N})+(1-t)T_{a}(t\mathcal{M}),
\end{align}
which is the convexity. (3) The strong monotonicity $T(\mathcal{N})\geq \sum_i q_i T({\mathcal{N}_i})$ can be proved by
\begin{align}
\sum_i q_i T_{a}({\mathcal{N}_i})&=\sum_i q_i \min_{\mathcal{F}_i\in\text{MIO}}\vert\vert({\mathcal{N}_i}-\mathcal{F}_i)\mathrm{\Delta}\vert\vert_1 \\
&=\sum_i q_i \vert\vert({\mathcal{N}_i}-\mathcal{F}_{i}^{*})\mathrm{\Delta}\vert\vert_1\\
&\leq \sum_i q_i  \min_{\mathcal{X}\in\text{MIO}}\vert\vert(\mathcal{F}_i-\mathcal{F}_i(X))\mathrm{\Delta}\vert\vert_1\label{submul}\\
&\leq \min_{\mathcal{X}\in\text{MIO}} \vert\vert(\mathcal{N}-X)\mathrm{\Delta}\vert\vert_1=T_{a}(\mathcal{N})\label{minsum},
\end{align}
 in which $\mathcal{N}_i=\mathfrak{F}_i(N)$ and $\mathfrak{F}_i$ are Kraus operators of superoperation $\mathfrak{F}$, $\mathcal{F}^{*}$  represents the channel  achieving the minimum, inequality (\ref{submul}) holds for that $\mathcal{F}_i(X)$ couldn't be the optimal, and inequality (\ref{minsum}) is valid due to the sub-multiplicativity.
  (4) The strong monotonicity combined with convexity leads to monotonicity. $\Box$

$T_{a,non}$ is a success probability in channel discrimination tasks if the participant allows the specific free dephasing operation or the incoherent states. This numerical result is very close to the one studied in Ref. \cite{qo}, where the authors analyzed detection-incoherent settings. Since the MIO is the maximal free set in static QRTs of coherence, constraints (such as applying a dephasing channel) would definitely shrink the free set (to dephasing incoherent channels).

\section{Dynamical total coherence in the RNI framework}

Quantum total coherence is one type of basis-independent coherence of which the static QRT was studied in Ref. \cite{YANG2018305}. Similarly, it can also be investigated in the sense of dynamical resource theory, i.e., the dynamical total coherence of a channel. In Ref.  \cite{YANG2018305}, it is explicitly given that the free operations with post-selective measurements are the mixed unitary channels defined as $\mathcal{U}(\cdot)=\sum_{x} q_x U_x(\cdot)U^{\dagger}_{x}$ with unitary $U$ and $\sum_x q_x=1$, and the free operations without post-selective measurements are the unital channels given by $\mathcal{A}(\cdot)=\sum_{x}  A_x(\cdot)A^{\dagger}_{x}$ with $\sum_{x} A_x(\cdot)A^{\dagger}_{x} =I$. Following the above section, for a given quantum channel $\mathcal{N}$ one can straightforwardly get the corresponding measures of the dynamical total coherence by replacing the set 'IO/MIO' in Theorem 6 and replacing the static coherence measure $\mathcal{M}$ by a proper total coherence measure. One can also easily show that these measures satisfy the necessary conditions for a valid dynamic resource theory.

Considering the computability, we'd like to mention that the static total coherence measure based on the $l_2$ norm is a good measure. In this sense, an explicit measure of the dynamical total coherence can be raised similar to theorem 7 as follows.

\textit{Theorem 9.-} For a quantum channel $\mathcal{N}$ with Kraus operators $\{K_n\}$, the dynamical total coherence in the RNI framework can be quantified  as
\begin{align}
\tilde{T}_{l_2}(\{K_n\})=\max\{\max_{\rho} \sum_{n} p_n \mathrm{Tr}[\rho_n^2]-\mathrm{Tr}[\rho^{2}],0\}\label{l2sep}
\end{align}
with post-selective measurements, and 
\begin{align}
\tilde{T}_{l_2}(\mathcal{N})=\max\{\max_{\rho}\mathrm{Tr}[ \sum_{n} p_n \rho_n]^2-\mathrm{Tr}[\rho^{2}],0\}\label{l2nonsep}
\end{align}
without post-selective measurements, where $\rho_n=\frac{K_n\rho K^{\dagger}_n}{p_n}$ and $p_n=\mathrm{Tr}(K_n \rho K_n^{\dagger})$.  

\textit{Proof.-}Since the static total coherence based on $l_2$ norm is a qualified measure (i.e. satisfy all constraints for measures including the monotonicity and the strong monotonicity) \cite{2016Total}, this theorem can entirely be understood as an explicit example of theorem 7. $\Box$

As demonstrations, we consider the dynamical coherence of amplitude damping channels which characterize the energy dissipation in the quantum process. With a given dissipation rate $\eta$, the Kraus operators of this channel can be given as  \cite{nielsen} 
\begin{align}
&K_{0}=
\left[
\begin{array} {lr}
1 & 0\\
0&\sqrt{1-\eta}\\
\end{array}
\right],
&K_{1}=
\left[
\begin{array} {lr}
0& \sqrt{\eta}\\
0&0\\
\end{array}
\right].
\end{align}
Given a density matrix in Bloch representaion
\begin{align}
&\rho=\frac{1}{2}
\left[
\begin{array} {lr}
1+z & x-iy\\
x+iy&1-z\\
\end{array}
\right],
\end{align}
where the parameters $x$, $y$ and $z$ are subject to $x^2+y^2+z^2\leq1$, the dynamical coherence measured by  $\tilde{T}_{l_2}(\{K_n\})$ with Eq. (\ref{l2sep}) reads \begin{align} \nonumber
\tilde{T}_{l_2}(\{K_n\})&=\max\{\max_{x,y,z}\frac{(1+z)^2+(1-\eta)^2(1-z)^2+2(1-\eta)(x^2+y^2)}{2(2-\eta+\eta z)}\\\label{functionf}
&+\frac{\eta(1-z)}{2}-\frac{1+x^2+y^2+z^2}{2}, 0\}
\end{align}
To maximize $\tilde{T}_{l_2}(\{K_n\}$, we would like to apply the method of \textit{residual Multipliers}:
 \begin{align}
\mathcal{L}(x, y, z, \lambda)=f(x, y, z)+\lambda(x^2+y^2+z^2-1),
 \end{align}
where  $f(x, y, z)$ is the main part (ignore the maximize) of Eq.(\ref{functionf}) and the residual part is the qubit constraint. The maximum of $f(x, y, z)$ can be obtained when the partial derivatives for all parameters equal zero and the inequality constraint holds for the equality. Thus, we conclude the following equations: 

\begin{equation}\label{Lasolve}
\left\{
\begin{array} {lr}
\displaystyle \frac{\partial{\mathcal{L}}}{\partial {x}}=\frac{2(1-\eta)x}{2-\eta+\eta z}-x+2\lambda x=0,\\
\displaystyle  \frac{\partial{\mathcal{L}}}{\partial {y}}=\frac{2(1-\eta)y}{2-\eta+\eta z}-y+2\lambda y=0,\\
\displaystyle \frac{\partial{\mathcal{L}}}{\partial {z}}=-2\eta z^3+(2\eta-6)z^2+(2\eta=4)z-2\eta+2=0,\\
\lambda(x^2+y^2+z^2-1)=0.
\end{array}
\right.
\end{equation}
It should be noted that for any $x^2+y^2+z^2=1$,  $f(x,y,z)=0$, which implies that the maximum lies within the feasible region. Thus, we have the optimal solution for Eq. (\ref{Lasolve}) as:
\begin{equation}
\left\{
\begin{array} {lr}
\lambda=x=y=0,\\
z=\frac{2 \eta+\sqrt{9-8 \eta}-3}{2 \eta}.
\end{array}
\right.
\end{equation}
Thus the analytic total coherence measure for amplitude damping channels is given by
\begin{equation}
\tilde{T}_{l_2}(\{K_n\})=\frac{9 \left(\sqrt{9-8 \eta}-3\right)-4 \eta \left(2 \eta+2 \sqrt{9-8 \eta}-9\right)}{4 \eta^2}.
\end{equation}
\begin{figure}
\centering
\includegraphics[scale=0.5]{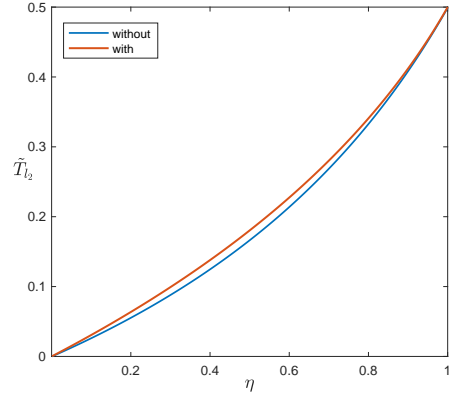}
\caption{The dynamical total coherence  by $\tilde{T}_2$  versus $\eta$.}\label{T12}
\end{figure}
For the situation without post-selective measurements, the dynamical total coherence coherence Eq. (\ref{l2nonsep}) is 
\begin{align}
\tilde{T}_{2}(\mathcal{N})=\mathrm{max}\{\sum\limits_{i=1}^3\frac{\xi_{i}^2\tilde{a}^2_{i}}{2(1-\xi^{2}_{i}))}+\frac{a^2_{i}}{2},\ 0\},
\end{align}
where   $a_{i}=\frac{1}{2}\mathrm{Tr}\mathcal{N}(\sigma_{i})$,  $M_{ij}=\frac{1}{2}\mathrm{Tr}\sigma_{i}\mathcal{N}(\sigma_{j})$, and  $\xi_{i}$ is the singular value of the matrix $M$,  $\vert\tilde{a}\rangle={U^{T}}\vert a \rangle$ with $U$ determined by the singular value decomposition ${M=U\Lambda V^{\mathrm{T}}}$ \cite{yang2021quantifying}. In Fig. \ref{T12} we compare the dynamical total coherence of amplitude damping channels with or without post-selective measurement. It is clear that the dynamical coherence with post-selective measurements is larger than that post-selective measurements.

\section{Conclusion and discussion}

Quantum channels are dynamical resources that contain more information than static states.  Investigating dynamical resources in QRTs has a pre-request that identifies the free sets in different physical backgrounds. In this paper, we introduce a framework in the sense of resource non-increasing. We introduce the free operation sets with and without post selective measurements.  It should be understood that the range of the RNI framework does not exceed the well-defined resource non-generating (RNG) framework, but it gives a new sight to determine the free set under different scenarios.  We design free super-operations with fundamental ingredients and give some measures to quantify dynamical resources. As a demonstration, we quantify dynamical (total) coherence in our frameworks. MIO and IO are free (incoherent) operations corresponding to the case without and with post selective measurements, respectively. The operational meanings in quantum tasks are also given for the dynamical coherence. In particular, the analytical calculation is given for the dynamical coherence of the amplitude damping channel, and the semidefinite programming is provided for dynamical coherence without post selective measurements.
 \section{Acknowledgments} This work was supported by the National Natural Science
Foundation of China under Grant No.12175029, No.11775040, and No. 12011530014.

\appendix

\section{Operational meaning of $T_{\diamond,non}$}

We first demonstrate the operational meaning for $T_{\diamond,non}$ in channel discrimination tasks \cite{discrimination2, qsl1,qsl2, qsl3}. In such a task, Bob wants to distinguish two channels with allowed operations. Another participant Alice prepares a probabilistic state in classic register Z. The classic register will be state 0 with probability $\lambda$ and state 1 with probability $1-\lambda$. Alice read out the state 0 or 1 in register Z, then sent an initial state (which is prepared in arbitrary register X) through quantum channels $\mathcal{N}_1$ and $\mathcal{N}_2$ respectively, and the final state ($\rho_1$ and $\rho_2$) was sent to Bob in register Y. Bob has to determine the final state had experienced which quantum channel. And Bob tries best to improve his successful probability by maximize $\frac{1}{2}\vert\vert\lambda\rho_{1}-(1-\lambda)\rho_{2}\vert\vert$.

We set $\lambda=\frac{1}{2}$  for a better illustration. Meanwhile, with practical consideration, Bob is allowed to hold an auxiliary register R and manipulate some operations $\mathrm{\Psi}^{\text{AR}}$. In a non-selective scenario, Bob can only rely on the information located in register Y.
Now, Bob, has a successful probability \cite{disprop,qo} as 

\begin{align}
\text{Prob}(\mathcal{N}_1,\mathcal{N}_{2})=\frac{1}{2}+\frac{1}{4}\displaystyle \max_{\mathrm{\Psi}^{\text{AR}},\sigma^{AR}}\vert\vert\mathrm{\Psi}^{\text{AR}}(\mathcal{N}_{1}-\mathcal{N}_{2})^{A}\otimes\mathds{1}^{\text{R}}\sigma^{\text{AR}}\vert\vert_1,\label{proba}
\end{align}
 where $\sigma^{AR}$ denotes the bipartite final state in registers Y and R. And the allowed manipulations  $\mathrm{\Psi}^{\text{AR}}$ can be determined by Bob if  are applied or not. If Bob holds the free set MIO and we minimize the diamond norm between resource channel $\mathcal{N}_1$ and  $\mathcal{N}_2\in\text{MIO}$, then we will find that
 \begin{align}\nonumber
 T_\diamond\left(\mathcal{N}_{1}\right)&=\displaystyle \min_{\mathcal{N}^{A}_2 \in \text{MIO}} \vert\vert\mathcal{N}^{A}_1-\mathcal{N}^{A}_2\vert\vert_\diamond\\\label{maxf}
& \geq \displaystyle \min_{\mathcal{N}^{A}_2 \in \text{MIO}}\displaystyle \max_{\mathcal{F}^{AR}\in \text{MIO}} \vert\vert\mathcal{F}^{AR}(\mathcal{N}_1-\mathcal{N}_2)^{A}\vert\vert_\diamond\\\nonumber
&=\displaystyle \min_{\mathcal{N}^{A}_2 \in \text{MIO}}\displaystyle \max_{\mathcal{F}^{AR}\in \text{MIO}} \vert\vert\mathcal{F}^{AR}(\mathcal{N}_1-\mathcal{N}_2)^{A}\otimes \mathds{1}^{R}\vert\vert_1\\\label{tra}
&=\displaystyle \min_{\mathcal{N}^{A}_2 \in \text{MIO}}\displaystyle \max_{\mathcal{F}^{AR}\in \text{MIO},P} \displaystyle \max_{\sigma^{AR}}\text{Tr}[P\mathcal{F}^{AR}(\mathcal{N}_1-\mathcal{N}_2)^{A}\otimes \mathds{1}^{AR}\sigma^{AR}]\\\nonumber
&=\displaystyle \min_{\mathcal{N}^{A}_2 \in \text{MIO}}\displaystyle \max_{\mathcal{F}^{AR}\in \text{MIO},\sigma^{AR}}\vert\vert\mathcal{F}(\mathcal{N}_1-\mathcal{N}_2)^{A}\otimes \mathds{1}^{AR}\sigma^{AR}\vert\vert_1\\\label{absorb}
\end{align}
where (\ref{maxf}) originates from the monotonicity of dynamical measures. (\ref{tra}) is an alternative definition for the trace norm $\left\Vert\cdot\right\Vert= \max_P \text{Tr} [P (\cdot)]$ where $P$ denotes  projective operators. Since any projective operators are in the set of  $\text{MIO}$, thus we have (\ref{absorb}).

Comparing (\ref{absorb}) with (\ref{proba}), one can easily see that in such a scenario the probability is exactly given by
 \begin{equation}
 \text{Prob}(\mathcal{N}_1,\mathcal{N}_{2})=\frac{1}{2}+\frac{1}{4}\displaystyle T_{\diamond, non}\left(\mathcal{N}_{1}\right),\label{hh}
 \end{equation}
since the maximum can be reached at least when Bob do an identity operation ( which belongs to MIO, i.e. a  free operation). Hence, the boundary gives the operational meaning to $T_{\diamond, non}$.

Furthermore, if Bob doesn't hold the ancilla reference R, one will directly derive an operational meaning for  $T_{1, non}$ based on Eq. (\ref{hh}).

\section{Semidefinite  programming for $T_{\diamond,non}$}

A quantum channel $\mathcal{N}$ has  its  Choi representation (sometimes called  Choi-Jamiolkowski isomorphism)\cite{CHOI,Jami} as
\begin{align}
J(\mathcal{N})=\mathds{1}\otimes\mathcal{N}(\phi_{+}),
\end{align}
where $\phi_{+}$ is unnormalized maximally entangled state $\phi_{+}=\sum_{ij}\vert i\rangle \langle j\vert\otimes\vert i\rangle \langle j\vert$. Its dynamical coherence can be measured by
\begin{align} 
T_{\diamond}(\mathcal{N})=\displaystyle \min_{\mathcal{F}\in \mathrm{FREE}}\vert\vert \mathcal{N}-\mathcal{F}\vert\vert_{\diamond}
\end{align} 
Under the RNI and non-selective background, the free set is MIO. Thus, the dynamical coherence measure  \begin{equation}T_{\diamond,non}(\mathcal{N})=\displaystyle \min_{\mathcal{F}\in\text{MIO}} \vert\vert\mathcal{N}-\mathcal{F}\vert\vert_{\diamond}
\end{equation}
and evaluated by semidefinite programming (with polynomial algorithms \cite{sdp1}). According to its definition, diamond norm for operator $\vert\vert \mathcal{N}-\mathcal{F}\vert\vert_{\diamond}$ has its primal problem
\begin{align}
&\textit{Primal}\\\nonumber
\mathrm{minimize} \quad &2\vert\vert\mathrm{Tr_{B}(Z)}\vert\vert_{\infty}\\\nonumber
\textit{s.t.}\quad &\mathrm{Z}\geq J(\mathcal{N}-\mathcal{F})\\\nonumber
&\mathrm{Z}\geq 0.
\end{align}
Then $T_{\diamond,non}(\mathcal{N})$ is the optimal value of 
\begin{align}
\mathrm{minimize}:\quad &2\vert\vert\mathrm{Tr_{B}(Z)}\vert\vert_{\infty}\\\nonumber
\textit{s.t.}\quad \mathrm{Z}&\geq J(\mathcal{N}-\mathcal{F})\\\nonumber
\mathrm{Z}&\geq 0\\\nonumber
\mathcal{F}&\in\:\mathrm{MIO}.
\end{align}

Applying constraints on $\mathcal{F}$ and Choi representation properties for  MIO, the primal problem becomes:
\begin{align}
\mathrm{minimize}:\quad &2\vert\vert\mathrm{Tr(Z)}\vert\vert_{\infty}\\\nonumber
\textit{s.t.}\quad \mathrm{Z}&\geq J(\mathcal{N})-M\\\nonumber
&\mathrm{Z}\geq 0\\\nonumber
&\mathrm{M}\geq0\\\nonumber
&\mathrm{Tr_{B}(M)}=\mathds{1}_{A}\\\nonumber
&\mathrm{Tr_{A}(M)}-\mathrm{\Delta}\mathrm{Tr_{A}(M)}=0.\nonumber
\end{align}
which equals to 
\begin{align}
\mathrm{minimize:}\: a\\\nonumber
\textit{s.t.} \quad &a\geq0\\\nonumber
&\mathds{1}_{A}\cdot a-2\mathrm{Tr_{B}(Z)}\geq 0\\\nonumber
& \mathrm{Z}\geq J(\mathcal{N})-M\\\nonumber
&\mathrm{Z}\geq 0\\\nonumber
&\mathrm{M}\geq0\\\nonumber
&\mathrm{Tr_{B}(M)}=\mathds{1}_{A}\\\nonumber
&\mathrm{Tr_{A}(M)}-\mathrm{\Delta}\mathrm{Tr_{A}(M)}=0.\nonumber
\end{align}
The Lagrangian of the primal problem is given by: 
\begin{align}
\mathcal{L}(a,Z,M,\widetilde{X},X,Y_1,Y_2)=a+\mathrm{Tr}[(2\mathrm{Tr_{B}(Z)}-a\mathds{1}_{A})\widetilde{X}]\\\nonumber
+\mathrm{Tr}[(J(\mathcal{N})-M-Z)X]+\mathrm{Tr}[(\mathrm{Tr_{A}M}-\mathrm{\Delta Tr_{A}M})Y_{1}]\\\nonumber
+\mathrm{Tr}[(\mathrm{Tr_{B}M}-\mathds{1}_{A})Y_{2}]
\end{align}
and its dual function (see more details for prime and dual problems in Ref. \cite{norms1}) is 
\begin{align*}
&q(\widetilde{X},X,Y_{1},Y_{2})=\displaystyle \inf_{a,Z,M } \mathcal{L}(a,Z,M,\widetilde{X},X,Y_1,Y_2)\\\nonumber
&\qquad=\displaystyle \inf_{a,Z,M } \mathrm{Tr}[J(\mathcal{N})]-\mathrm{Tr}[Y_2]+a[1-\mathrm{Tr}(\widetilde{X})]\\\nonumber
&+\mathrm{Tr}[(2\cdot\widetilde{X}\otimes\mathds{1}_{B}-X)Z]+\mathrm{Tr}[(\mathds{1}_{A}\otimes Y_{1}-\mathds{1}_{A}\otimes \mathrm{\Delta}Y_{1}+Y_{2}\otimes \mathds{1}_{B}-X)M].\nonumber
\end{align*}
The dual function has value $\mathrm{Tr}[J(\mathcal{N})X]-\mathrm{Tr}[Y_2]$ if $ \widetilde{X}\leq 1\land 2\cdot\widetilde{X}\otimes\mathds{1}_{B}-X\geq0\:\land\mathds{1}_{A}\otimes Y_{1} -\mathds{1}\otimes\mathrm{\Delta}Y_{1}+Y_{2}\otimes\mathds{1}_{B}-X\geq0$ and  $-\infty$ in other cases. 

Thus, the dual problem is to maximize $\mathrm{Tr}[J(\mathcal{N})X]-\mathrm{Tr}[Y_2]$  with constraints for $\widetilde{X}, X, Y_1,Y_2$. To make the constraint  clear and easy reading, we simplify the four constraints: (1-2) $\widetilde{X}$ is nonnegative  and trace less than one; (3) $X \geq 0$;  (4) $2\cdot\widetilde{X}\otimes\mathds{1}_{B}-X$ to one constraint  as $X\leq 2\cdot\rho\otimes\mathds{1}_{B}$ which $ \rho$  is a density matrix . Then we can construct a 
$\widetilde{X}^{'}:=\frac{1}{\mathrm{Tr}\widetilde{X}}\widetilde{X}$ is trace one,  positive semidefinite and will keep $2\cdot\widetilde{X}^{'}\otimes\mathds{1}_{B}-X$ positive semidefinite for all X satisfy $2\widetilde{X}\otimes\mathds{1}_{B}-X\geq 0$.  Hence, the dual problem is 
\begin{align}
&Dual \\\nonumber
\mathrm{maximize}\quad &\mathrm{Tr}[J(\mathcal{N})X]-\mathrm{Tr}[Y_{2}]\\\nonumber
s.t.\quad& X\leq 2\cdot\rho\otimes\mathds{1}_{B}: \rho\mathrm{\:is\:density\: matrix}\\\nonumber
&\mathds{1}_{A}\otimes Y_{1}-\mathds{1}\otimes \mathrm{\Delta}Y_{1}+Y_{2}\otimes\mathds{1}_{B}-X\geq0\\\nonumber
&X\geq0\\\nonumber
&Y_{1}=Y_{1}^{\dagger}\\\nonumber
&Y_{2}=Y_{2}^{\dagger}.\nonumber
\end{align}

In the end, we have to show that the primal and the dual problem reaches the same optimal value, that is the strong duality holds in this programming.
The strong duality obtains if the Slater condition \cite{slater} holds: there exists some $Z^{*}$ and $W^{*}$ satisfy all the equality constraints in primal problem and  all the inequality constraints \textit{strictly} hold. It is easy to find that the Slater condition holds when  
\begin{align}
Z^{*}&=\mathds{1}_{A}\otimes\mathds{1}_{B}+J(\mathcal{N})\\\nonumber
M^{*}&=\frac{1}{ \vert B\vert}\mathds{1}_{A}\otimes\mathds{1}_{B}.  
\end{align}
and the strong duality would promise the optimal value can be reached.

\textit{Example.} The quantum channel $\mathcal{K}$ has two Kraus operators:
\begin{align}
&K_{0}=
\left[
\begin{array} {lr}
0.2096&   -0.3956\\
-0.2564&  -0.3719\\
\end{array}
\right],\\
&K_{1}=
\left[
\begin{array} {lr}
-0.6197 &   0.6418\\
 -0.7116&  -0.5415\\
\end{array}
\right].
\end{align}
 By solving SDP in software CVX \cite{cvx}, this quantum channel has total coherence as +0.186758.

\bibliographystyle{iopart-num}%
\bibliography{ref}

\providecommand{\noopsort}[1]{}\providecommand{\singleletter}[1]{#1}%
\providecommand{\newblock}{}
\begin{thebibliography}{10}
\expandafter\ifx\csname url\endcsname\relax
  \def\url#1{{\tt #1}}\fi
\expandafter\ifx\csname urlprefix\endcsname\relax\def\urlprefix{URL }\fi
\providecommand{\eprint}[2][]{\url{#2}}

\bibitem{entanglement}
Horodecki R, Horodecki P, Horodecki M and Horodecki K 2009 {\em Rev. Mod.
  Phys.\/} {\bf 81}(2) 865--942
  \urlprefix\url{https://link.aps.org/doi/10.1103/RevModPhys.81.865}

\bibitem{entangle}
Vedral V, Plenio M~B, Rippin M~A and Knight P~L 1997 {\em Phys. Rev. Lett.\/}
  {\bf 78}(12) 2275--2279
  \urlprefix\url{https://link.aps.org/doi/10.1103/PhysRevLett.78.2275}

\bibitem{e5}
Yu C~S and Song H~S 2009 {\em Phys. Rev. A\/} {\bf 80}(2) 022324
  \urlprefix\url{https://link.aps.org/doi/10.1103/PhysRevA.80.022324}

\bibitem{discord}
Ollivier H and Zurek W~H 2001 {\em Phys. Rev. Lett.\/} {\bf 88}(1) 017901
  \urlprefix\url{https://link.aps.org/doi/10.1103/PhysRevLett.88.017901}

\bibitem{qd1}
Datta A, Shaji A and Caves C~M 2008 {\em Phys. Rev. Lett.\/} {\bf 100}(5)
  050502
  \urlprefix\url{https://link.aps.org/doi/10.1103/PhysRevLett.100.050502}

\bibitem{qd2}
Piani M, Cavalcanti D, Aolita L, Boixo S, Modi K and Winter A 2011 {\em APS
  Meeting Abstracts\/}

\bibitem{qd3}
Luo S 2008 {\em Phys. Rev. A\/} {\bf 77}(4) 042303
  \urlprefix\url{https://link.aps.org/doi/10.1103/PhysRevA.77.042303}

\bibitem{qd4}
Giorda P and Paris M~G~A 2010 {\em Phys. Rev. Lett.\/} {\bf 105}(2) 020503
  \urlprefix\url{https://link.aps.org/doi/10.1103/PhysRevLett.105.020503}

\bibitem{Quantifying}
Baumgratz T, Cramer M and Plenio M~B 2014 {\em Phys. Rev. Lett.\/} {\bf
  113}(14) 140401
  \urlprefix\url{https://link.aps.org/doi/10.1103/PhysRevLett.113.140401}

\bibitem{nonlocality}
Brunner N, Cavalcanti D, Pironio S, Scarani V and Wehner S 2014 {\em Rev. Mod.
  Phys.\/} {\bf 86}(2) 419--478
  \urlprefix\url{https://link.aps.org/doi/10.1103/RevModPhys.86.419}

\bibitem{contextuality}
Kirchmair G, Z{\"a}hringer F, Gerritsma R, Kleinmann M, G{\"u}hne O, Cabello A,
  Blatt R and Roos C~F 2009 {\em Nature\/} {\bf 460} 494--497
  \urlprefix\url{https://doi.org/10.1038/nature08172}

\bibitem{Strobel424}
Strobel H, Muessel W, Linnemann D, Zibold T, Hume D~B, Pezz{\`e} L, Smerzi A
  and Oberthaler M~K 2014 {\em Science\/} {\bf 345} 424--427 ISSN 0036-8075
  \urlprefix\url{https://science.sciencemag.org/content/345/6195/424}

\bibitem{as}
Marvian I and Spekkens R 2014 {\em Nat. Commun.\/} {\bf 5} 3821

\bibitem{alter}
Yu X~D, Zhang D~J, Xu G~F and Tong D~M 2016 {\em Phys. Rev. A\/} {\bf 94}(6)
  060302(R) \urlprefix\url{https://link.aps.org/doi/10.1103/PhysRevA.94.060302}

\bibitem{trace}
Rana S, Parashar P and Lewenstein M 2016 {\em Phys. Rev. A\/} {\bf 93}(1)
  012110 \urlprefix\url{https://link.aps.org/doi/10.1103/PhysRevA.93.012110}

\bibitem{multi}
Yao Y, Xiao X, Ge L and Sun C~P 2015 {\em Phys. Rev. A\/} {\bf 92}(2) 022112
  \urlprefix\url{https://link.aps.org/doi/10.1103/PhysRevA.92.022112}

\bibitem{mea1}
Zhao H~Q and Yu C~S 2018 {\em Sci. Rep.\/} {\bf 8} 299

\bibitem{mea2}
Yu C~S 2017 {\em Phys. Rev. A\/} {\bf 95}(4) 042337
  \urlprefix\url{https://link.aps.org/doi/10.1103/PhysRevA.95.042337}

\bibitem{mre}
Bu K, Singh U, Fei S~M, Pati A~K and Wu J 2017 {\em Phys. Rev. Lett.\/} {\bf
  119}(15) 150405
  \urlprefix\url{https://link.aps.org/doi/10.1103/PhysRevLett.119.150405}

\bibitem{mea3}
Wu Z, Zhang L, Fei S~M and Li-Jost X 2020 {\em J. Phys. A: Math. Theor.\/} {\bf
  54} 015302 \urlprefix\url{https://doi.org/10.1088/1751-8121/abcab7}

\bibitem{op1}
Winter A and Yang D 2016 {\em Phys. Rev. Lett.\/} {\bf 116}(12) 120404
  \urlprefix\url{https://link.aps.org/doi/10.1103/PhysRevLett.116.120404}

\bibitem{op2}
Napoli C, Bromley T~R, Cianciaruso M, Piani M, Johnston N and Adesso G 2016
  {\em Phys. Rev. Lett.\/} {\bf 116}(15) 150502
  \urlprefix\url{https://link.aps.org/doi/10.1103/PhysRevLett.116.150502}

\bibitem{op3}
Rana S, Parashar P, Winter A and Lewenstein M 2017 {\em Phys. Rev. A\/} {\bf
  96}(5) 052336
  \urlprefix\url{https://link.aps.org/doi/10.1103/PhysRevA.96.052336}

\bibitem{op4}
Zhu H, Hayashi M and Chen L 2018 {\em Phys. Rev. A\/} {\bf 97}(2) 022342
  \urlprefix\url{https://link.aps.org/doi/10.1103/PhysRevA.97.022342}

\bibitem{op6}
Patel D, Patro S, Vanarasa C, Chakrabarty I and Pati A~K 2021 {\em Phys. Rev.
  A\/} {\bf 103}(2) 022422
  \urlprefix\url{https://link.aps.org/doi/10.1103/PhysRevA.103.022422}

\bibitem{ass1}
Marvian I, Spekkens R~W and Zanardi P 2016 {\em Phys. Rev. A\/} {\bf 93}(5)
  052331 \urlprefix\url{https://link.aps.org/doi/10.1103/PhysRevA.93.052331}

\bibitem{piani}
Piani M, Cianciaruso M, Bromley T~R, Napoli C, Johnston N and Adesso G 2016
  {\em Phys. Rev. A\/} {\bf 93}(4) 042107
  \urlprefix\url{https://link.aps.org/doi/10.1103/PhysRevA.93.042107}

\bibitem{as3}
Ioffe L and M\'ezard M 2007 {\em Phys. Rev. A\/} {\bf 75}(3) 032345
  \urlprefix\url{https://link.aps.org/doi/10.1103/PhysRevA.75.032345}

\bibitem{ch3}
Korzekwa K, Czach{\'{o}}rski S, Pucha{\l}a Z and {\.{Z}}yczkowski K 2018 {\em
  New J. Phys.\/} {\bf 20} 043028
  \urlprefix\url{https://doi.org/10.1088/1367-2630/aaaff3}

\bibitem{ch5}
Wang X, Wilde M~M and Su Y 2019 {\em New J. Phys.\/} {\bf 21} 103002
  \urlprefix\url{https://doi.org/10.1088/1367-2630/ab451d}

\bibitem{POVM1}
Bischof F, Kampermann H and Bru\ss{} D 2019 {\em Phys. Rev. Lett.\/} {\bf
  123}(11) 110402
  \urlprefix\url{https://link.aps.org/doi/10.1103/PhysRevLett.123.110402}

\bibitem{qrt5}
Saxena G, Chitambar E and Gour G 2020 {\em Phys. Rev. Research\/} {\bf 2}(2)
  023298
  \urlprefix\url{https://link.aps.org/doi/10.1103/PhysRevResearch.2.023298}

\bibitem{ch9}
Xu J 2021 {\em Phys. Lett. A\/} {\bf 387} 127028 ISSN 0375-9601
  \urlprefix\url{https://www.sciencedirect.com/science/article/pii/S0375960120308951}

\bibitem{ch8}
Gour G and Scandolo C~M 2020 {\em Phys. Rev. Lett.\/} {\bf 125}(18) 180505
  \urlprefix\url{https://link.aps.org/doi/10.1103/PhysRevLett.125.180505}

\bibitem{ch1}
Theurer T, Satyajit S and Plenio M~B 2020 {\em Phys. Rev. Lett.\/} {\bf
  125}(13) 130401
  \urlprefix\url{https://link.aps.org/doi/10.1103/PhysRevLett.125.130401}

\bibitem{qrt1}
Chitambar E and Gour G 2019 {\em Rev. Mod. Phys.\/} {\bf 91}(2) 025001
  \urlprefix\url{https://link.aps.org/doi/10.1103/RevModPhys.91.025001}

\bibitem{nch1}
Liu Y and Yuan X 2020 {\em Phys. Rev. Research\/} {\bf 2}(1) 012035(R)
  \urlprefix\url{https://link.aps.org/doi/10.1103/PhysRevResearch.2.012035}

\bibitem{nch2}
Kuroiwa K and Yamasaki H 2020 {\em {Quantum}\/} {\bf 4} 355 ISSN 2521-327X
  \urlprefix\url{https://doi.org/10.22331/q-2020-11-01-355}

\bibitem{nch3}
D{\'{i}}az M~G, Desef B, Rosati M, Egloff D, Calsamiglia J, Smirne A,
  Skotiniotis M and Huelga S~F 2020 {\em {Quantum}\/} {\bf 4} 249 ISSN
  2521-327X \urlprefix\url{https://doi.org/10.22331/q-2020-04-02-249}

\bibitem{nch4}
Designolle S, Uola R, Luoma K and Brunner N 2021 {\em Phys. Rev. Lett.\/} {\bf
  126}(22) 220404
  \urlprefix\url{https://link.aps.org/doi/10.1103/PhysRevLett.126.220404}

\bibitem{qd}
Gour G and Winter A 2019 {\em Phys. Rev. Lett.\/} {\bf 123}(15) 150401
  \urlprefix\url{https://link.aps.org/doi/10.1103/PhysRevLett.123.150401}

\bibitem{nch5}
Masini M, Theurer T and Plenio M~B 2021 {\em Phys. Rev. A\/} {\bf 103}(4)
  042426 \urlprefix\url{https://link.aps.org/doi/10.1103/PhysRevA.103.042426}

\bibitem{nch6}
Hsieh C~Y 2021 {\em PRX Quantum\/} {\bf 2}(2) 020318
  \urlprefix\url{https://link.aps.org/doi/10.1103/PRXQuantum.2.020318}

\bibitem{qo}
Theurer T, Egloff D, Zhang L and Plenio M~B 2019 {\em Phys. Rev. Lett.\/} {\bf
  122}(19) 190405
  \urlprefix\url{https://link.aps.org/doi/10.1103/PhysRevLett.122.190405}

\bibitem{PhysRevA.92.032331}
Mani A and Karimipour V 2015 {\em Phys. Rev. A\/} {\bf 92}(3) 032331
  \urlprefix\url{https://link.aps.org/doi/10.1103/PhysRevA.92.032331}

\bibitem{YANG2018305}
Yang S~R and Yu C~S 2018 {\em Ann. Phys.\/} {\bf 388} 305--314 ISSN 0003-4916
  \urlprefix\url{https://www.sciencedirect.com/science/article/pii/S0003491617303500}

\bibitem{2016Total}
Yu C~S, Yang S~R and Guo B~Q 2016 {\em Quantum Inf. Process\/} {\bf 15}
  3773--3784

\bibitem{nielsen}
Nielsen M~A and Chuang I 2000 {\em Quantum computation and quantum
  information\/} (Cambridge University Press)

\bibitem{yang2021quantifying}
ren Yang S and shui Yu C 2021 Quantifying dynamical total coherence in a
  resource non-increasing framework (\textit{Preprint} \eprint{2110.14267})

\bibitem{discrimination2}
Chiribella G, D'Ariano G~M and Perinotti P 2008 {\em Phys. Rev. Lett.\/} {\bf
  101}(18) 180501
  \urlprefix\url{https://link.aps.org/doi/10.1103/PhysRevLett.101.180501}

\bibitem{qsl1}
Cooney T, Mosonyi M and Wilde M~M 2016 {\em Commun. Math. Phys.\/} {\bf 344}
  797--829 \urlprefix\url{https://doi.org/10.1007/s00220-016-2645-4}

\bibitem{qsl2}
Hayashi M 2009 {\em IEEE Trans. Inf. Theory\/} {\bf 55} 3807--3820

\bibitem{qsl3}
Duan R, Feng Y and Ying M 2009 {\em Phys. Rev. Lett.\/} {\bf 103}(21) 210501
  \urlprefix\url{https://link.aps.org/doi/10.1103/PhysRevLett.103.210501}

\bibitem{disprop}
Matthews W, Wehner S and Winter A 2009 {\em Communications in Mathematical
  Physics\/} {\bf 291} 813--843

\bibitem{CHOI}
Choi M~D 1975 {\em Linear Algebra Appl.\/} {\bf 10} 285--290 ISSN 0024-3795
  \urlprefix\url{https://www.sciencedirect.com/science/article/pii/0024379575900750}

\bibitem{Jami}
Jamio{\l}kowski A 1972 {\em Rep. Math. Phys.\/} {\bf 3} 275--278 ISSN 0034-4877
  \urlprefix\url{https://www.sciencedirect.com/science/article/pii/0034487772900110}

\bibitem{sdp1}
Khachiyan L 1980 {\em USSR Comput. Math. Phys.\/} {\bf 20} 53--72 ISSN
  0041-5553
  \urlprefix\url{https://www.sciencedirect.com/science/article/pii/0041555380900610}

\bibitem{norms1}
Watrous J 2018 {\em The theory of quantum information\/} (Cambridge university
  press)

\bibitem{slater}
Giorgi G and Kjeldsen T~H 2014 {\em Traces and emergence of nonlinear
  programming\/} (Springer)

\bibitem{cvx}
Grant M and Boyd S Cvx: Matlab software for disciplined convex programming
  \url{http:///cvxr.com}

\end{thebibliography}

\end{document}